\documentstyle[prb,aps]{revtex}
\begin{document}
\draft
\title{Vertical magneto-tunneling through a quantum dot and the density 
 of states of small electronic systems}
\author{Augusto Gonzalez\cite{augusto}$^{1,2}$ and
 Roberto Capote\cite{capote}$^3$}
\address{$^1$ Departamento de Fisica, Universidad de Antioquia, AA 1226,
 Medellin, Colombia\\
 $^2$Instituto de Cibernetica, Matematica y Fisica Calle E 
 309, Vedado, Habana 4, Cuba\\ 
 $^3$Centro de Estudios Aplicados al Desarrollo Nuclear,
 Calle 30 No 502, Miramar,\\ La Habana, Cuba}
\date{\today}

\maketitle

\begin{abstract}
One-electron tunneling through a quantum dot with a 
strong magnetic field in the direction 
of the current is studied. The linear magneto-conductance
is computed for a model parabolic dot with seven electrons in the
intermediate states and for different values of the magnetic field. 
It is shown that the dot density of states at low excitation energies 
can be extracted from a precise measurement of the conductance at the 
upper edge of the Coulomb blockade diamond.
We parametrized the density of states with a single ``temperature''
parameter (in the so called ``constant temperature approximation''), and 
found that this parameter depends very weakly on the magnetic field.
\end{abstract}


\section{Introduction}

The experimental study of vertical transport through a semiconductor
quantum dot \cite{Tarucha98} have stressed the similarities of these small 
electronic systems with real atoms. Shell effects (in a confinement 
potential which is approximately parabolic in shape) and spin effects are
clearly distinguished in the positions of conductance peaks \cite{Tarucha96}.
The inclusion of a relatively strong magnetic field, $B$, makes possible
to measure spin reordering with $B$ up to the formation of a completely
polarized electronic droplet \cite{Oosterkamp99}. The technique has also 
revealed its power as a spectroscopic tool in the determination of the
low-lying energy levels of the few-electron dots \cite{Spectroscopy}.
More recent developments include integer spin Kondo effects 
\cite{Sasaki2000}, and the study of two-electron tunneling in the Coulomb
blockade regime \cite{DeFranceschi2000}.

The position of conductance peaks, obtained from the experiments, determine 
addition energies and even excitation energies to the first excited states,
if they are well separated from the rest of the states. This only happens
for the few-electron dots and for the very first excited states. In the 
six-electron dot, for example, at excitation energies around 1 meV, the 
density of states may be as high as 80 levels/meV. That is, mean level 
distance around 0.012 meV (see Section III). In the present paper, we show
that for relatively small dots and excitation energies $\le$ 1 meV, the
density of states can be obtained from a precise measurement of the height
and position of the conductance peak at the upper edges of the Coulomb 
blockade diamonds.

We present model calculations for a 6-electron dot in a magnetic field
$8.75\le B\le 12$ T. Our model parameters are chosen to approximate the
experimental conditions in papers
\onlinecite{Oosterkamp99,DeFranceschi2000}. The relatively strong magnetic 
field guarantees that
only spin-polarized states are relevant in tunneling processes. On the 
other hand, the temperature is low enough (a few mK) for thermal excitation
to be neglected. Thus a pure quantum mechanical description is used.

The plan of the paper is as follows. The details of the model quantum dot
are specified in Section II. In the next section, the low-lying energy 
levels of the 6- and 7-electron dots are computed, and the density of
levels at low excitation energies is parametrized with a single 
``temperature'' parameter. We show 
that this parameter depends very weakly on the magnetic field. The 
calculation of the linear magneto-conductance (transmission coefficient) is
sketched in Section IV. In Section V, we show how the conductance at 
maxima can be 
related to the density of energy levels, and we compare the estimate
following from the conductance with the actual (calculated) density of
states. 

\section{The model quantum dot}

A schematic representation of the vertical profile ($z$ axis) of the 
bottom of the conduction band for the model structure to be used is 
given in Fig. \ref{fig1}a. This is a symmetric 
AlGaAs(7 nm)-InGaAs(12 nm)-AlGaAs(7 nm) quantum well, with n-doped
source (S) and drain (D) contacts \cite{DeFranceschi2000}. The quantum dot 
is formed within the well region. For the smallest dots, the lateral ($xy$)
confinement is approximately parabolic, with $\hbar\omega_0=3$ meV
\cite{Oosterkamp99}.

The source potential will be taken as the reference potential. It will be
fixed to zero. 
The barrier height due to the Al concentration is estimated in Ref.
\onlinecite{DeFranceschi2000} as 50 meV. We will add 42 meV corresponding
to the Coulomb barrier (6 electrons times 7 meV). Thus, the top of the 
left barrier is at 92 meV, and for the right barrier at (92-$V_{sd}$) meV 
($V_g$ and $V_{sd}$ are in fact energies, not voltages). The drain 
potential is $-V_{sd}$.
Notice that our $V_g$ parameter is not the one used in experiments. In 
our model, the depth of the potential well is $-(V_g+V_{sd}/2)$, while
in experiments $V_g$ is related to the depth of the well through the
capacitance of the system.

The electron mass and dielectric constant are taken as $0.067 m_0$ and 12.5
respectively all over the structure. In the computation of the $N$-electron
states in the dot, a first quantum well sub-band approximation is used.
The energy of the first quantum well state, $E_z$, is computed numerically
as a function of $V_g$ and $V_{sd}$. The total energy is then written

\begin{equation}
E=N E_z+E_{xy}, 
\end{equation}

\noindent
where $E_{xy}$ is obtained from a two-dimensional calculation which 
includes the lateral confinement and the effective Coulomb potential
(equal to 0.8 times the two-dimensional potential to account for averaging 
in the $z$ direction).

Magnetic field values around 10 T are considered. In this region, the 
electronic droplet is completely spin polarized \cite{Oosterkamp99}. 
Moreover, the separation between Landau levels (LL) is 
$\hbar\omega_c=1.728\; B$ meV, where $B$ is measured in Teslas. Thus, the 
second LL is around 17 meV above the first. If we are interested in 
excitation energies below 1meV, we may neglect contributions from the
second and higher LL's \cite{Serna2000}. In section III, a large basis of 
Slater first LL functions is used to construct the two-dimensional matrix 
hamiltonian, which is further diagonalized by means of a Lanczos algorithm.

In section IV, we will compute the transmission coefficient of the model
shown in Fig. \ref{fig1}, in which the potential is sectionally constant,
unlike a real structure.

\section{Density of levels at low excitation energies}

In the present section, we show results for the spin-polarized energy 
levels of the 6- and 7-electron dots. The starting point is the 
``first LL'' $H_{xy}$ hamiltonian 

\begin{equation}
H_{xy}=(|M|+N)\hbar\Omega -|M|\hbar\omega_c/2+\beta \sum_{i,j,k,l}
 <i,j|1/r|k,l>a_i^+ a_j^+ a_l a_k, 
\label{hamiltoniano}
\end{equation}

\noindent
where $\omega_c=e B/(m c)$ is the cyclotronic frequency, and $\Omega=
\sqrt{\omega_0^2+(\hbar\omega_c/2)^2}$. The sum runs over indexes, $l$,
representing angular momentum projection onto the $z$ axis. $a_l^+$ 
creates one electron in a harmonic oscillator state of frequency $\Omega$
with zero radial quantum number and $l\le 0$. The Zeeman energy is not 
included. The constant $\beta$ equals $0.8\; e^2/(\kappa l_{\Omega})$, 
where $l_{\Omega}$ is the harmonic oscillator length, and $\kappa$ -- the
dielectric constant.

The hamiltonian $H_{xy}$ is diagonalized in a basis of Slater determinants 
with fixed angular momentum projection $M=\sum_{t=1}^N l_t$. For a given 
$M$, this basis is finite. In a 7-electron system, for example, the sector
with $M=-80$ contains 40340 functions. These large matrices are better 
diagonalized with a Lanczos algorithm.

We show in Fig. \ref{fig2} the lowest energy levels 
(excitation energy $\le$ 1 meV) for the $N=6$ and 7-electron dots in
magnetic fields $B=8.75$  and 12 T. A few remarkable facts are evident 
from this figure. The average number of levels, $n$, with excitation 
energies $\le \Delta E$ may be very well fitted by a ``constant 
temperature approximation'' \cite{T0}

\begin{equation}
n=\exp (\Delta E/\Theta).
\label{eq3}
\end{equation}

The temperature parameter, $\Theta$, exhibits a weak dependence on the 
magnetic field. When $B$ varies from 8.75 to 12 T, for example, $1/\Theta$
for the 6-electron dot changes only from 2.99 to 3.20 (meV)$^{-1}$. In the
language of ``filling factors'', $\nu\approx M_0/M_{gs}$ (where $M_0=
-N(N-1)/2$ is the momentum corresponding to filling factor one), one has
$\nu=15/35=3/7$ at $B=8.75$ T, and $\nu=15/45=1/3$ for $B=12$ T. The gap 
to the first excited state shrinks to zero at the $B$ values where the 
ground-state momentum, $M_{gs}$, changes first from -35 to -39, and then
from -39 to -45. $\Theta$, however, varies very weakly. It means that $\Theta$ 
is not a measure of this gap, but of the actual low-lying density of levels.

The average density of levels following from Eq. (\ref{eq3}) is

\begin{equation}
\frac{{\rm d}n}{{\rm d} E}=\frac{1}{\Theta} \exp (\Delta E/\Theta), 
\end{equation} 

\noindent
which may be taken as a smooth version of the actual density

\begin{equation}
\frac{{\rm d}n}{{\rm d} E}=\sum_r \delta(\Delta E-\Delta E_r), 
\end{equation} 

\noindent
where $\Delta E_r$ denotes the excitation energy of the $r$-th level.
For $N=6$ and $B=12$ T, for example, the average density is around 80 
levels/meV at $\Delta E$=1 meV.

\section{Linear magneto-conductance}

We will compute the conductance, ${\rm d}I/{\rm d}V_{sd}$, from the 
transmission coefficient, $T$, by means of a simplified Landauer expression
\cite{Ando98}

\begin{equation}
\frac{{\rm d}I}{{\rm d}V_{sd}}=\frac{e^2}{h} T. 
\label{eq6}
\end{equation}

\noindent
Only one spin polarization is considered in (\ref{eq6}) due to the
quenching of the spin-down states by the magnetic field. $T$ is computed
from the usual relation between transmitted and incident current in the
$z$ direction.

The fact that there is only one electron tunneling through the structure
allows one to write an ansatz for the total wave function in which $N$
electrons are permanently confined inside the dot. We restrict our
attention to 6 confined, and a seventh tunneling electron. Thus, we shall
work in the interval of $V_g$ values corresponding to the N=6 Coulomb
blockade diamond of Fig. \ref{fig1}b. We shall focus on the first
conductance peak when $V_{sd}$ is varied, i. e. the edge of the diamond.

As mentioned above, only spin polarized states are considered. The reason 
is the following. Initial states for the tunneling processes are states 
with $N$ electrons in the dot and one free electron in the source electrode.
The energy is given in Eq. (\ref{eq9}). Contributions to the trasmission 
coefficient come from intermediate states of $N+1$ electrons in the dot,
and final states with $N$ electrons in the dot and one electron in the drain,
which energy is less or equal than the energy given in Eq. (\ref{eq9}). As
we are interested in the first conductance peak when $V_{sd}$ is increased,
only one intermediate state contributes, i. e. the ground state of $N+1$
electrons in the dot. In a strong magnetic field, this is a spin polarized
state. The long spin-relaxation times in quantum dots due to phase space
reduction \cite{vina99,kn2000}, make spin-flip processes in the final 
stage of tunneling impossible. Thus, relevant final states are also spin
polarized.

The total wave function is written in a separable way: $\Psi=\Psi_z
\Psi_{xy}$. The ansatz for the $z$ function is the following:

\begin{equation}
\Psi_z=\chi(z_7) \prod_{u=1}^6 \chi_1(z_u),
\end{equation}  

\noindent
where $\chi_1$ is the first quantum well function. We will determine the
combination $\chi(z_7) \Psi_{xy}$ in each $z$ interval where the
$z$-potential is sectionally constant. For $\Psi_{xy}$, a Fock
representation will be used. We will simplify the notation further,
writing $|\alpha\rangle$ instead of $\Psi_{xy}(N)$, and $|\gamma\rangle$
instead of $\Psi_{xy}(N+1)$. $|\alpha_0\rangle$ will denote the ground
state. Then, the ansatz for the wave function is the following

(1) $z<0$:

\begin{equation}
\Psi_1=\hat a_{l_0}^{\dagger} |\alpha_0\rangle \left\{\exp{i k_1 z}+b_1
\exp{-i k_1 z} \right\},
\end{equation} 

\noindent
where $k_1=\sqrt{2 m \epsilon_1}/\hbar$, and the total energy is written
(apart from trivial constants) as

\begin{equation}
E=E_{\alpha_0}+\frac{\hbar\omega_c}{2}+\epsilon_1.
\label{eq9}
\end{equation}

\noindent
$\epsilon_1$ is the initial kinetic energy of the tunneling electron. We
will fix it at a small value, $\epsilon_1=0.01$ meV. $\hat
a_{l_0}^\dagger$ is the $(xy)$ creation operator for an electron with
angular momentum projection $l_0$ in S. Due to the assumed cylindrical
symmetry, the total angular momentum, $M=M_{\alpha_0}+l_0$ is a conserved
quantity.

(2) $0<z<L_B$:

\begin{equation}
\Psi_2=\hat a_{l_0}^{\dagger} |\alpha_0\rangle \left\{a_2\exp{-k_2 z}+b_2
\exp{k_2 z} \right\},
\end{equation} 

\noindent
where $\epsilon_2=\epsilon_1$, and  $k_2=\sqrt{2 m
(V_B-\epsilon_2)}/\hbar$. $V_B=92$ meV is the barrier height. 

(3) $L_B<z<L_B+L$:

\begin{equation}
\Psi_3=\sum_\gamma |\gamma\rangle \left\{a_3^\gamma \exp{i k_3^\gamma
z}+b_3^\gamma \exp{-i k_3^\gamma z} \right\},
\end{equation} 

\noindent
where $k_3^\gamma=\sqrt{2 m
(V_g+V_{sd}/2+\epsilon_3^\gamma)}/\hbar$, and
$\epsilon_3^\gamma=\epsilon_1+E_{\alpha_0}+\hbar\omega_c/2-E_{\gamma}$. 

(4) $L_B+L<z<2 L_B+L$:

\begin{equation}
\Psi_4=\sum_\alpha \hat a_{l_\alpha}^\dagger|\alpha\rangle
\left\{a_4^\alpha \exp{-k_4^\alpha z}+b_4^\alpha \exp{k_4^\alpha z}
\right\},
\end{equation} 

\noindent
where $k_4^\alpha=\sqrt{2 m
(V_B-V_{sd}-\epsilon_4^\alpha)}/\hbar$, and
$\epsilon_4^\alpha=\epsilon_1+E_{\alpha_0}-E_{\alpha}$. Finally, 

(5) $2 L_B+L<z$:

\begin{equation}
\Psi_5=\sum_\alpha \hat a_{l_\alpha}^\dagger|\alpha\rangle
a_5^\alpha \exp{i k_5^\alpha z},
\label{eq13}
\end{equation} 

\noindent
where $\epsilon_5^\alpha=\epsilon_4^{\alpha}$, and  
$k_5^\alpha=\sqrt{2 m(V_{sd}+\epsilon_5^\alpha)}/\hbar$. The sum in
(\ref{eq13}) runs over open final state channels, i. e.

\begin{equation}
E_\alpha<\epsilon_1+E_{\alpha_0}+V_{sd}.
\end{equation}

In the above formulae, $L_B=7$ nm and $L=12$ nm are the barrier and well
widths respectively. Notice that our ansatz for $\Psi$ respects a weak
version of the Pauli principle, but it is not completely antisymmetrized
with respect to the seventh electron.

The continuity of the current leads to relations among the coefficients
$a$ and $b$ in the different regions. With our ansatz, however, we found
impossible to satisfy the continuity relations for general linear
combinations of intermediate
or final states. It means that we shall compute the
transmission coefficient for each pair $(\gamma,\alpha)$ independently. We
will refer to the pair $(\gamma,\alpha)$ as a tunneling channel. We will
be particularly interested in the $(\gamma_0,\alpha)$ channels, where the
intermediate state is the ground state of $N+1$ electrons in the dot. The
upper edge of the $N=6$ Coulomb blockade diamond in Fig. \ref{fig1}b is
related to these channels. Notice that the absence of interference between
tunneling channels should be further reinforced by temperature effects.
Note also that overlapping coefficients of transversal ($x, y$) functions
cancel out in ratios $b/a$, and thus in $T$.  

The partial transmission coefficient, $T_\alpha$ is defined as usual

\begin{equation}
T_\alpha=I_D(\alpha)/I_S,
\end{equation} 

\noindent
where $I_S$ is the incident current, and $I_D(\alpha)$ is the transmitted
current when only the $\alpha$ channel is considered open. The total
transmission coefficient is obtained from charge conservation arguments. If
there are a few open channels, then $I_S T_1$ is the current flowing
through channel 1, $I_S (1-T_1) T_2$ is the current through channel 2,
etc. The total coefficient is then

\begin{equation}
T=1-\prod_{open\;channels}(1-T_\alpha).
\label{eq16}
\end{equation} 
 
\section{Results and discussion}

We show in Fig. \ref{fig3}a the partial transmission coefficient,
$T_2$ corresponding to the excited state $\alpha=2$ at excitation
energy $\Delta E_2=0.425$ meV in a magnetic field $B=12$ T. This
channel is closed for $V_{sd}<0.42$ meV, as expected. Three situations are
depicted: below resonance ($V_g=39.89$ meV), maximum resonance
($V_g=39.87$ meV), and above resonance ($V_g=39.6$ meV). The asymptotic
shape of the curve below resonance is typical. In Nuclear Physics context,
it is interpreted as interference between resonance and potential
scattering \cite{nuclphys}. The maximum resonance occurs at a $V_{sd}$
slightly higher than $\Delta E_2$. Note that, as the incident
electron energy is fixed, the maximum of $T_2$ is not one, but a
value around 0.85. Above resonance, the $\alpha=2$ channel remains open 
but the transmission coefficient diminishes. 

The computation of $T$ from Eq. (\ref{eq16}) show results like the one
drawn in Fig. \ref{fig3}b, where all of the open channels 
$(\gamma,\alpha)$ at a given
$V_{sd}$ are included. $V_g$ is 39.89 meV, and $B=12$ T. The distance
between the two peaks, 0.6 meV, is approximately twice the excitation
energy to the first excited state of $N+1$ electrons in the dot. Below, we
will focus only on the first peak. 

The first two (ground and
first excited) $\alpha$ states contribute to this peak. Both are above
resonance. The second and higher excited states are below resonance. Their
contribution to $T$ is very little. Thus, there are not channels at
maximum resonance, and the value of $T$ at the peak is lower than the
maximum $T_2^{res}\sim 0.85$ for the $\alpha=2$ channel. With a small
decrease of $V_g$, $T$ reaches the value $T_2^{res}$, meaning that there
is one channel at maximum resonance. Notice that the distance between the
first and second excited $\alpha$ states is only 0.12 meV. It means that
only levels which are very close in energy may be simultaneously
resonant. The width of the one-channel peak, $\sim 0.05$ meV, may serve as
an estimation of the resonance interval.

The sensitivity of the first peak maximum with the number of levels in the
resonance interval may be used for an experimental estimation of the
density of levels at low excitation energies. We prove this statement in
Fig. \ref{fig4}. The number of levels in an energy interval $\delta=0.05$
meV below the excitation energy $\Delta E$ is drawn along with the 
estimate (points) obtained from the conductance in the following way.  

We follow the first peak maximum for different $V_g$. Its position,
i. e. the value of $V_{sd}$, is identified with the excitation energy
$\Delta E$, and from the height, $T_{peak}$, we obtain the number of
levels in the resonance interval as

\begin{equation}
n=\ln (1-T_{peak})/\ln (1-T_1^{res}),
\label{eq17}
\end{equation} 

\noindent
where $T_1^{res}$ is the value of $T$ when the first excited state
$\alpha=1$ is at maximum resonance. The idea behind Eq. (\ref{eq17}) is
that the product in Eq. (\ref{eq16}) may be approximated as
$(1-T_1^{res})^n$.

We compare in Figs. \ref{fig4} (a) and (b) the estimation coming from
Eq. (\ref{eq17}) with the exact level distribution obtained before. The
agreement is excellent. Abrupt variations of the density of levels are
nicely reproduced as well as the smooth behaviour obtained from
Eq. (\ref{eq3}).  
This agreement proves the factibility of measuring the low-energy density
of levels in few-electron quantum dots by means of a precise measurement
of the conductance.

In conclusion, we have computed the density of spin-polarized levels of 6-
and 7-electron quantum dots at low excitation energies and strong
magnetic fields. The density is well parametrized by a constant
temperature approximation. The temperature parameter, $\Theta$, shows a
weak dependence on the magnetic field. We computed also the dot
conductance (transmission coefficient) for vertical tunneling, and showed
that the conductance at the upper edge of the Coulomb blockade diamond is
directly related to the density of levels. In this way, a procedure is
suggested for the experimental determination of the low-lying density of
states from a precise measurement of the conductance. Altough calculations
were carried out under the simplifying assumptions of zero temperature and
strong magnetic fields, extensions to zero or weak fields and finite
temperatures are also possible.

\acknowledgements 
A. G. acknowledges the Research Committee of the University of 
Antioquia (CODI) for support.

\begin{figure}
\caption{(a) Schematic conduction band profile for the quantum dot under 
study. The magnetic field is aligned with the current along the $z$ axis.
(b) The N=6 Coulomb blockade diamond. We compute the conductance for a
fixed $V_g$ and varying $V_{sd}$, as represented by the dashed line.}
\label{fig1}
\end{figure}

\begin{figure}
\caption{The low-lying energy levels of the $N=6$ and $N=7$ dots at  
$B=8.75$  and 12 T.}
\label{fig2}
\end{figure}

\begin{figure}
\caption{(a) Partial transmission coefficient corresponding to the second 
excited (final) state at $B=12$ T. (b) Total transmission at $V_g=39.89$
meV and $B=12$ T.}
\label{fig3}
\end{figure}

\begin{figure}
\caption{Number of levels in the resonance interval $(\Delta
E-\delta,\Delta E)$ with $\delta=0.05$ meV. Exact results and the 
estimate coming from Eq. (\ref{eq17}) are compared. 
(a) $B=12$ T. (b) $B=8.75$ T.}
\label{fig4}
\end{figure}

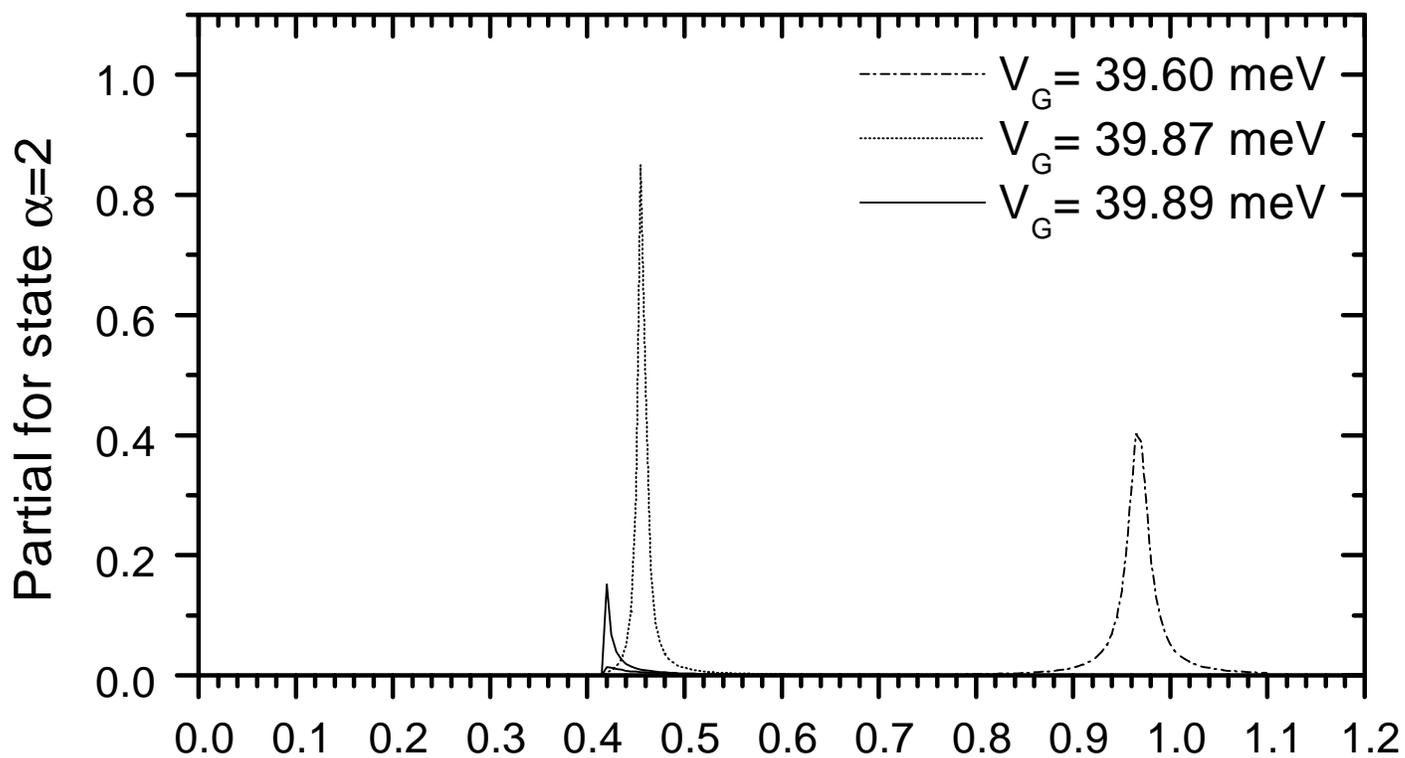
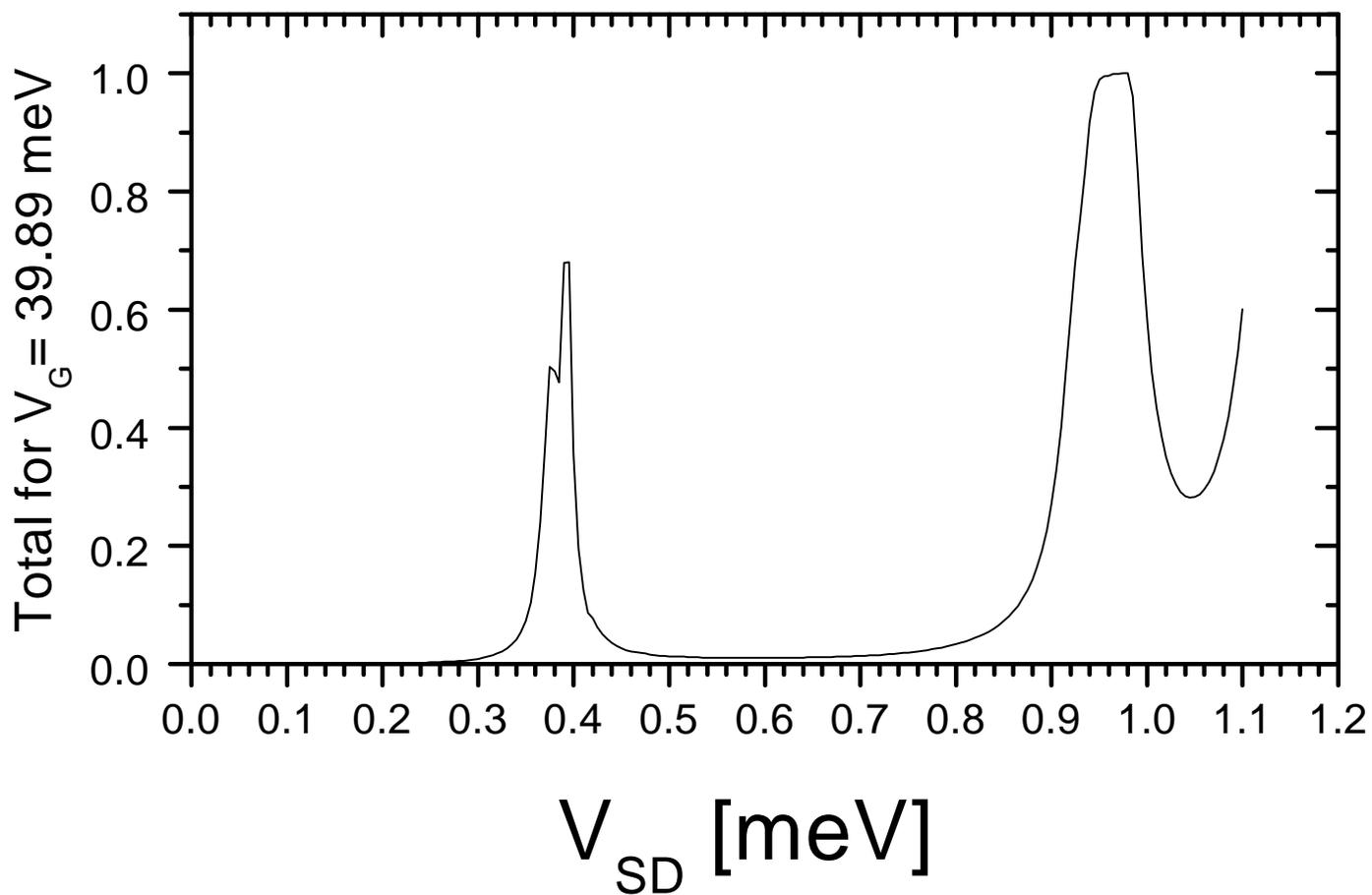
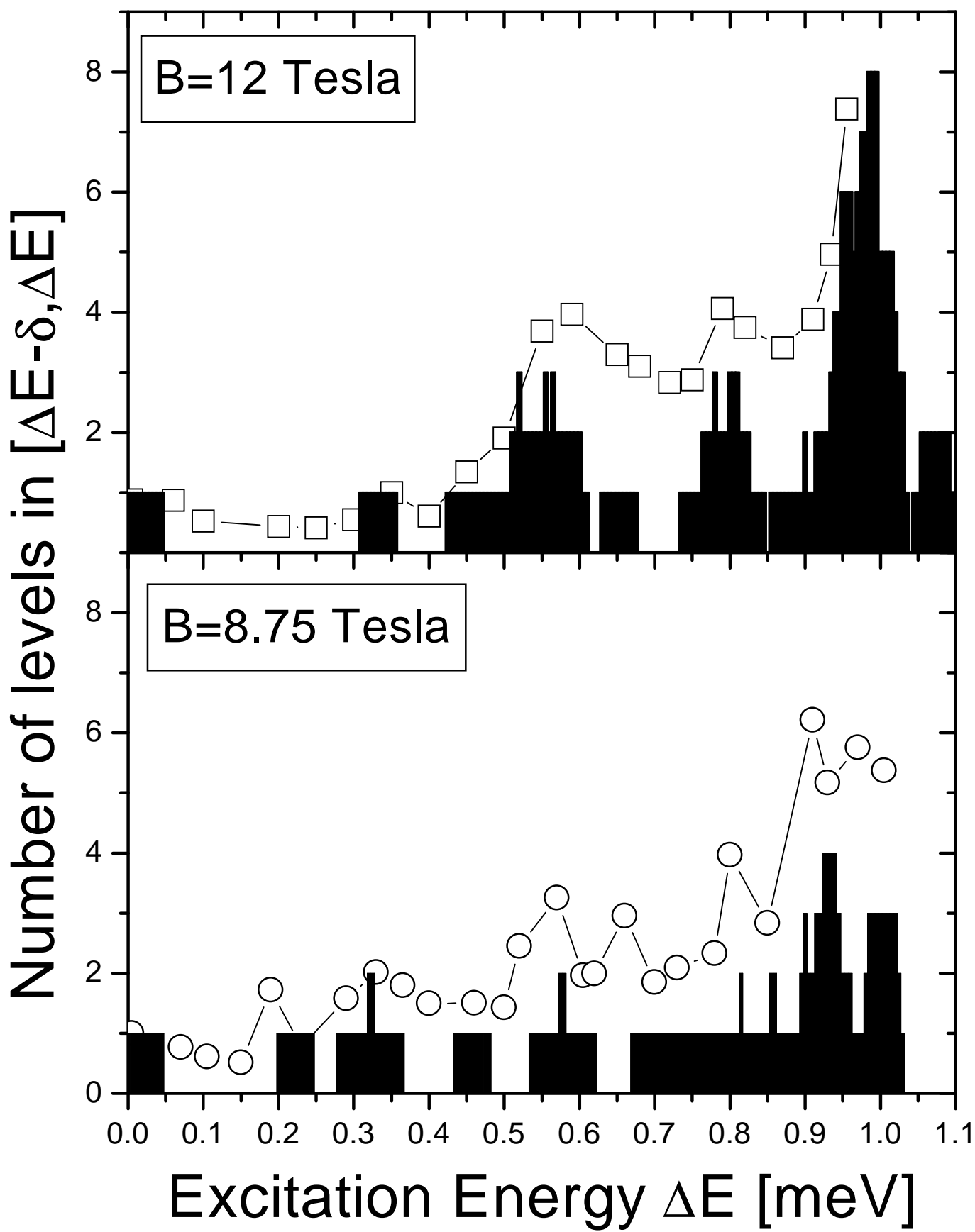
\end{document}